\documentclass[a4paper]{jpconf}
\usepackage{graphicx}
\usepackage{citesort}

\begin{document}
\title{Fully microscopic shell-model calculations with realistic
  effective hamiltonians}

\author{L. Coraggio$^1$, A. Covello$^{1,2}$, A.Gargano$^1$,
  N. Itaco$^{1,2}$, and T. T. S. Kuo$^3$}

\address{$^1$Istituto Nazionale di Fisica Nucleare and $^2$
  Dipartimento di Scienze Fisiche, Universit\`a di Napoli Federico II, \\
Complesso Universitario di Monte  S. Angelo, Via Cintia - I-80126 Napoli,
Italy\\
$^3$Department of Physics, SUNY, Stony Brook, New York 11794, USA}

\ead{luigi.coraggio@na.infn.it}

\begin{abstract}
The advent of nucleon-nucleon potentials derived from chiral
perturbation theory, as well as the so-called $V_{\rm low-k}$ approach
to the renormalization of the strong short-range repulsion contained
in the potentials, have brought renewed interest in realistic
shell-model calculations.
Here we focus on calculations where a fully microscopic approach is
adopted.
No phenomenological input is needed in these calculations, because
single-particle energies, matrix elements of the two-body interaction,
and matrix elements of the electromagnetic multipole operators are
derived theoretically.
This has been done within the framework of the time-dependent
degenerate linked-diagram perturbation theory.
We present results for some nuclei in different mass regions. These
evidence the ability of realistic effective hamiltonians to provide an
accurate description of nuclear structure properties.
\end{abstract}

\section{Introduction}
The main problem with nuclear shell-model calculations is the
choice of the input ingredients, namely the single-particle (SP)
energies and the matrix elements of the residual two-body interaction
(TBME).

A phenomenological approach to this problem, which has been
successfully applied in different mass regions, is to treat these
quantities as parameters to be least-squares fitted to a chosen set of
experimental data (see for instance \cite{ABrown01}).

A more fundamental approach is to start from the free nucleon-nucleon
($NN$) potential and derive an effective shell-model hamiltonian in
the framework of a many-body theory (see for instance
\cite{Coraggio09a,Hjorth95}).
Usually, the SP energies are taken from experiment and only the
TBME are retained from the theoretical effective hamiltonian.
This approach, that started with the pioneering work of Kuo and Brown
\cite{Kuo66} for the $sd$-shell region, in the last decade  has proven
to be a viable way to provide an accurate description of nuclear
structure properties. 
A review of some results may be found in Ref. \cite{Coraggio09a}.

However, the advances in computational power and the
refinement of the time-dependent degenerate linked-diagram
perturbation theory \cite{Coraggio09a} to derive the shell-model
effective hamiltonian have paved the way to a fully
microscopic approach to the nuclear shell-model, that is to employ
both SP energies and TBME derived from theory.

Here, we present some examples of fully-microscopic shell-model
calculations we have performed in different mass regions
\cite{Coraggio05c,Coraggio07b,Coraggio09d,Coraggio10a}, to verify the
practical value of this approach. 

In the following section, we outline the perturbative approach to the
derivation of a realistic shell-model hamiltonian.
In Section III, we present results of shell-model calculations
for heavy carbon and oxygen isotopes, $pf$-shell nuclei, and
$N=82$ isotones.
A short summary is given in the last section.

\section{Outline of calculations}
We calculate the two-body shell-model effective hamiltonian within the
framework of the time-dependent degenerate linked-diagram expansion
\cite{Coraggio09a}.

This means that we describe the wave function of a nucleus with two
valence nucleons as a two-nucleon state whose energy is calculated
taking into account the interaction of the two valence nucleons with
the closed core perturbatively.
To this end, an auxiliary one-body potential $U$ is introduced to
break up the hamiltonian as the sum of an unperturbed term $H_0$, 
which describes the independent motion of the nucleons, and a residual 
interaction $H_1$:

\[
H=\sum_{i=1}^{A} \frac{p_i^2}{2m} + \sum_{i<j} V^{ij}_{NN} = T + V_{NN} =
(T+U)+(V_{NN}-U)= H_{0}+H_{1}~~.
\]

The effective hamiltonian $H_{\rm eff}$ is expressed through the
Kuo-Lee-Ratcliff folded-diagram expansion in terms of the vertex
function $\hat{Q}$-box, which is composed of irreducible valence-linked
diagrams \cite{Coraggio09a}.
We include in the $\hat{Q}$-box one- and two-body Goldstone diagrams
through third order in $H_1$.
The folded-diagram series is summed up to all orders using the
Lee-Suzuki iteration method \cite{Suzuki80}.

$H_{\rm eff}$ contains one-body contributions, whose collection is the
so-called $\hat{S}$-box \cite{Shurpin83}.
In realistic shell-model calculations it is customary to use a
subtraction procedure, so that only the two-body terms of $H_{\rm
  eff}$, which make up the effective interaction $V_{\rm eff}$, are
retained while the SP energies are taken from experiment
\cite{Covello01}.
In fully microscopic shell-model calculations, we adopt a different
approach employing SP energies obtained from the $\hat{S}$-box
calculation.
In this regard, it is worth pointing out that, owing to the presence
of the $-U$ term in $H_1$, self-consistency correction diagrams arise
in the $\hat{Q}$-box.
In our calculation we use the harmonic oscillator potential, and take
into account all self-consistency correction diagrams up to third
order.

Let us now focus our attention on the input interaction $V_{NN}$. 
It would of course be very desirable to use directly a realistic $NN$
potential that reproduces the two-body scattering data and deuteron
properties with high precision.
However, to perform nuclear structure calculations with realistic $NN$
potentials in a perturbative framework, one has first to deal with
the strong repulsive behavior of such potentials in the high-momentum 
regime.
An advantageous method to renormalize the bare $NN$ interaction 
has been proposed in \cite{Bogner02}.
It consists in deriving from $V_{NN}$ a low-momentum potential 
$V_{\rm low-k}$ defined within a cutoff momentum $\Lambda$ by way of a
similarity transformation. 
This is a smooth potential which preserves exactly the onshell
properties of the original $V_{NN}$ and is suitable for being used 
directly in nuclear structure calculations \cite{Coraggio09a}.

We have derived from the high-precision CD-Bonn $NN$ potential
\cite{Machleidt01b} a $V_{\rm low-k}$ corresponding to a value
of the cutoff $\Lambda=2.6$ fm$^{-1}$, and this potential as been
employed to derive shell-model effective hamiltonians both for
$pf$-shell nuclei and $N=82$ isotones.

A new and alternative approach to the renormalization of $NN$
potential with a repulsive short-range component, such as the CD-Bonn
one, is to employ a low-momentum realistic interaction derived from
chiral perturbation theory at next-to-next-to-next-to-leading order.
This is the so-called N$^3$LOW $NN$ potential
\cite{Coraggio07b} with a sharp momentum cutoff at 2.1 fm$^{-1}$,
which we have employed to derive shell-model hamiltonians for heavy
carbon and oxygen isotopes.

\section{Results and comparison with experiment}
In this section, we show and discuss results of shell-model
calculations we have performed for nuclei with mass ranging from
$A=16$ to $A=154$.
Let us start with the isotopic chain of the heavy carbon isotopes for
which we have used an effective shell-model hamiltonian derived from
the N$^3$LOW potential, with $^{14}$C considered as an inert core
\cite{Coraggio10a}.

\begin{figure}[h]
\begin{minipage}{18pc}
\includegraphics[width=9pc,angle=90]{Fig1.epsi}
\caption{\label{Fig1} (Color online) Experimental
  \cite{Audi03,Tanaka10short} and calculated ground-state  energies
  for carbon isotopes from $A=16$ to 24. $N$ is the number of
  neutrons. See text for details.}
\end{minipage}\hspace{2pc}%
\begin{minipage}{18pc}
\begin{center}
\includegraphics[width=9pc,angle=90]{Fig2.epsi}
\caption{\label{Fig2} (Color online) Experimental
  \cite{Ong08short,Elekes09short} and calculated excitation energies
  of the yrast $J^{\pi}=2^+$ states for carbon isotopes from $A=16$ to
  22. $N$ is the number of neutrons.}
\end{center}
\end{minipage} 
\end{figure}

This is a challenging subject, because in a recent Letter
\cite{Tanaka10short} Tanaka {\it et al.} have reported on the
observation of the neutron dripline for the carbon isotopes, which has
been located at $^{22}$C.
This nucleus may be considered the heaviest Borromean nucleus ever observed
\cite{Kemper10} and it may be seen as composed of three parts: two
neutrons plus $^{20}$C.
These three pieces must all be present in order to obtain a bound
nucleus, as is proved by the particle instability of $^{21}$C.

In Fig. \ref{Fig1} the calculated ground-state (g.s.) energies of the
even-mass isotopes (continuous black line) relative to $^{14}$C are
compared with the experimental ones (continuous red line)
\cite{Audi03}.
The experimental behavior is well reproduced, in particular our
results confirm that $^{22}$C is the last bound isotope; its
calculated two-neutron separation energy $S_{2n}$ is 601 keV to be
compared with the evaluation of 420 keV \cite{Audi03}.
Moreover, our calculations predict that $^{21}$C is unstable against
one-neutron decay, the theoretical $S_n$ being -1.6 MeV.
Therefore, our results fit the picture of $^{22}$C as a Borromean
nucleus.

In Fig. \ref{Fig2} we report the experimental excitation energies of
the yrast $2^+$ states as a function of $N$ and compare them
with our calculated values.
It can be seen that the observed energies are nicely reproduced.
We also report our predicted excitation energy, 4.661 MeV, for the
unbound $J^{\pi}=2^+_1$ state in $^{22}$C.

\begin{figure}[h]
\begin{minipage}{18pc}
\includegraphics[width=9pc,angle=90]{Fig3.epsi}
\caption{\label{Fig3}(Color online)  Experimental
  \cite{Audi03} and calculated ground-state  energies for oxygen
  isotopes from $A=18$ to 28. $N$ is the number of neutrons. The red
  dashed line refers to (non-experimental) estimated values
  \cite{Audi03}. See text for details.}
\end{minipage}\hspace{2pc}%
\begin{minipage}{18pc}
\begin{center}
\includegraphics[width=9pc,angle=90]{Fig4.epsi}
\caption{\label{Fig4}(Color online) Experimental \cite{nndc} and
  calculated excitation energies of the yrast $J^{\pi}=2^+$ states for
  oxygen isotopes from $A=18$ to 24. $N$ is the number of neutrons.}
\end{center}
\end{minipage} 
\end{figure}

A similar calculation, starting also from the N$^3$LOW potential, has
been performed for the chain of the oxygen isotopes, where $^{16}$O
has been considered as the inert core.

In Fig. \ref{Fig3} the calculated g.s. energies of
even-mass isotopes (continuous black line) relative to $^{16}$O are
compared with the experimental ones (continuous red line)
\cite{Audi03}.
From the inspection of Fig. \ref{Fig3}, it can be seen that our
calculations overestimate the experimental data.
It is worth noting that this discrepancy may be ``healed'' by
upshifting the calculated SP spectrum so as to reproduce
the experimental g.s. energy of $^{17}$O relative to $^{16}$O.
The results obtained with this upshift ($427$ keV) are reported in
Fig. \ref{Fig3} by the black dashed line, and we see that the neutron
dripline is located at $^{24}$O.
The need for this shift can be probably traced to the lack in our
hamiltonian of a three-body force in addition to the N$^3$LOW two-body
potential \cite{Coraggio07b}.

In Fig. \ref{Fig4} the experimental and calculated excitation energies
of the yrast $2^+$ states are reported as a function of $A$.
It can be seen that the observed energies, as well as the subshell
closures at $N=14$ and 16, are nicely reproduced.
It should be noted that experimentally the $N=14$ subshell closure
disappears when moving from oxygen to carbon isotopes
\cite{Hoffman09short}.
We see that this is correctly reproduced by our calculations when
comparing the results shown in Figs. \ref{Fig2} and \ref{Fig4}.

As mentioned in Section II, we have also performed calculations
starting from a $V_{\rm low-k}$ derived from the CD-Bonn potential.
Let us now examine the region of $pf$-shell nuclei which have been
studied deriving an effective hamiltonian for valence nucleons
interacting in the four $pf$ orbitals outside doubly-closed
$^{40}$Ca.

In Fig. \ref{Fig5} we have plotted the calculated (continuous black line)
and experimental (continuous red line) g.s. energies per
valence neutron of even-mass calcium isotopes, relative to $^{40}$Ca,
as a function of $A$.
As in the case of the oxygen isotopes, our calculations overestimate
the experimental data, and this may again be traced to the lack of the
inclusion of a three-body force that could correct the calculated SP
spectrum. 
In fact, if we upshift the latter so as to reproduce the experimental
$^{41}$Ca binding energy respect to $^{40}$Ca, our results are
upshifted by about 1.6 MeV (black dashed line), thus leading to good
agreement with the experimental data along the whole isotopic chain.

\begin{figure}[h]
\begin{minipage}{18pc}
\includegraphics[width=9pc,angle=90]{Fig5.epsi}
\caption{\label{Fig5}(Color online)  Experimental \cite{Audi03} and calculated
  ground-state  energies per valence neutron for calcium isotopes from
  $A=42$ to 56. $N_{\rm val}$ is the number of valence neutrons. See
  text for details.}
\end{minipage}\hspace{2pc}%
\begin{minipage}{18pc}
\includegraphics[width=9pc,angle=90]{Fig6.epsi}
\caption{\label{Fig6}(Color online) Experimental \cite{nndc} and calculated
  excitation energies of the yrast $J^{\pi}=2^+$ states for calcium
  isotopes from $A=42$ to 54.}
\end{minipage} 
\end{figure}

The good quality of our effective interaction for calcium isotopes is
further confirmed by the inspection of Fig. \ref{Fig6}, where the
experimental and calculated excitation energies of the yrast $2^+$
states are reported as a function of $A$.
It can be seen that our calculations reproduce nicely the observed
energies, apart form a quenching of the $N=28$ shell closure.
It should be  mentioned that in the calculations for Ca isotopes
reported in Ref. \cite{Coraggio09c}, we have employed an empirical set
of SP energies, whose main difference respect to the theoretical one is
the larger energy gap between the $p_{3/2}$ and $f_{7/2}$ levels, that
is mainly responsible for the excitation energy of $J^{\pi}=2^+$ state
in $^{48}$Ca.

\begin{figure}[ht]
\begin{center}
\includegraphics[width=14pc,angle=90]{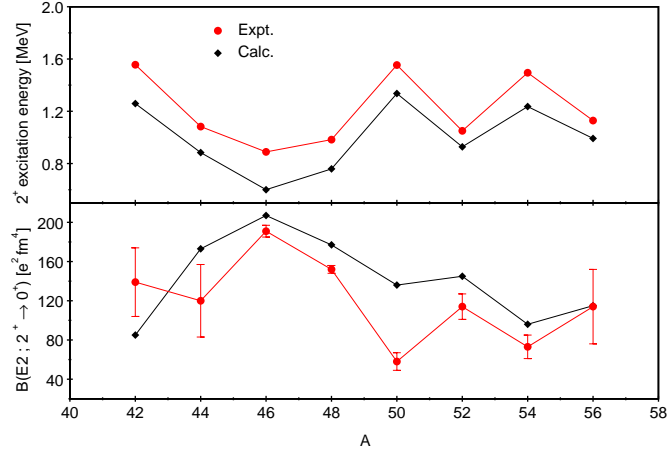}
\caption{\label{Fig7}(Color online) Experimental \cite{nndc} and
  calculated excitation energies of the yrast $J^{\pi}=2^+$ states and
  their relative $B(E2;2^+_1 \rightarrow 0^+_1)$  transition rates for
  titanium isotopes from $A=42$ to 56.}
\end{center}
\end{figure}

In order to prove the reliability of our calculations when dealing
with valence proton-neutron systems, in Fig. \ref{Fig7} we have
plotted the excitation energy of the yrast $2^+$ state as a function
of $A$ in even-mass Ti isotopes.
In the same figure we also report the $B(E2;2^+_1 \rightarrow 0^+_1)$
transition rates, calculated employing an effective operator obtained
at third order in perturbation theory, consistently with the
derivation of $H_{\rm eff}$.
A relevant result is that the theory reproduces the staggering of the
transition rates for the heavy isotopes.
Note that the enhancement of the $B(E2)$ transition rate in $^{56}$Ti
seems to suggest a softening of the $N=34$ subshell closure
\cite{Poves05}.

\begin{figure}[h]
\begin{minipage}{16pc}
\includegraphics[width=16pc]{Fig8.epsi}
\caption{\label{Fig8}(Color online) Experimental and calculated
  ground-state energies per valence proton for $N=82$ isotones from
  $A=134$ to 154. $Z_{\rm val}$ is the number of valence protons.}
\end{minipage}\hspace{6pc}%
\begin{minipage}{12pc}
\includegraphics[width=11pc,angle=0]{Fig9.epsi}
\caption{\label{Fig9} Experimental and calculated $^{134}$Te spectra.}
\end{minipage} 
\end{figure}

Finally, we present some selected results obtained for the even-mass
$N=82$ isotones starting from a $V_{\rm low-k}$ derived from the
CD-Bonn potential \cite{Coraggio09d}.

In Fig. \ref{Fig8}, we show the calculated and experimental
\cite{Audi03} g.s. energies (relative to the $^{132}$Sn core) per
valence proton of even-mass isotones as a function of the number of
valence particles $Z_{\rm val}$. 
We see that the experimental and theoretical curves are practically
straight lines having the same slope while being about 2.4 MeV apart.
This discrepancy is essentially the same as that existing between  the
theoretical and experimental g.s. energies of $^{133}$Sb.
This confirms the reliability of our SP spacings and TBME, since the
pattern of the theoretical curve depends only on these quantities.

A strong test for the effective hamiltonian is given by the
calculation of the energy spectrum of $^{134}$Te, since the theory of
the effective interaction is tailored for systems with two valence
nucleons.
In Fig. \ref{Fig9}, the experimental \cite{nndc} and calculated
$^{134}$Te spectra are reported up to 3.5 MeV excitation energy. It is
worth mentioning that the quality of the results is comparable to that
obtained in our previous calculations where the SP energies have been
taken from  experiment \cite{Coraggio08a}.

\section{Concluding remarks}
We have presented some selected results of shell-model studies within
the framework of a fully microscopic approach.
This means that, starting from a realistic $NN$ potential, an
effective hamiltonian has been derived by way of perturbation theory
and then employed in our shell-model calculations without use of any
adjustable parameter.

When compared with experiment, the quality of our results shows the
practical value of this approach, as they reproduce correctly
saturation and shell-closure properties, as well as many different
spectroscopic features of the nuclear structure.

In our opinion these calculations, that make use of both
single-particle energies and two-body matrix elements derived from the
theory, may provide reliable predictions that do not depend on any
phenomenological input.

\section*{References}
\bibliography{biblio}

\end{document}